\documentclass[aps,prl,twocolumn,showpacs,nofootinbib]{revtex4}

 \usepackage{amsmath,amssymb,bm,euscript}
 \usepackage{graphicx,boxedminipage}
 \usepackage{floatflt}
 \usepackage[hypertex]{hyperref}
 \usepackage{xcolor,fancybox}

\definecolor{Dark-blue}{RGB}{0,0,255}
\newcommand{\bra}[1]{\ensuremath{\bm{\langle}#1\bm{|}}}
\newcommand{\ket}[1]{\ensuremath{\bm{|}#1\bm{\rangle}}}

\begin{document}

\title{The mechanism of quantum chaos manifestations in the spectra of singularly perturbed wave-billiard systems}

\author{E.\,M. Ganapolskii}

\author{Yu.\,V. Tarasov}

 \affiliation{A.\,Ya.\,Usikov Institute for Radiophysics and Electronics, NAS of Ukraine,\\ 12 Proscura Street, 61085 Kharkov, Ukraine}
\begin{abstract}
  The spectra of a microwave cylindrical resonator with the embedded thin metal rod playing the role of a singular perturbation are studied both theoretically and experimentally. The intra- and inter-mode scattering caused by the perturbation are clearly distinguished and recognized to play essentially different parts in the appearance of spectrum chaotic properties. The analysis based on the mode-mixing operator norm shows that the inter-mode scattering dominates over the intra-mode scattering and basically determines statistical properties of the resonator spectrum. The results we have obtained in the experiment are in good conformity with our theory. Clear manifestations of quantum chaos are revealed for the resonator with the asymmetrically inserted rod, namely, the Wigner-type distribution of the inter-frequency intervals, the apparent correlation between spectral lines, and the characteristic curve of the spectral rigidity. By comparing the theory and the experiment we succeeded in establishing for the first time that it is just the inter-mode scattering that is responsible for the quantum chaos manifestations in the singularly perturbed integrable wave-billiard system.
\end{abstract}
\date{\today}
\maketitle

The present paper is focused on the study of quantum chaos in linear wave-resonance systems similar to classical Sinai billiard. Nowadays this problem attracts much attention, as is demonstrated by numerous theoretical and experimental researches in the field (see, e.\,g., Ref.~\cite{bib:Stockman99} and references therein). The concept of quantum chaos (QC) is quite general, spanning the broad range of researches related to quantum-mechanical description of systems possessing chaotic properties in the classical limit. It is for this reason that the studies of QC are of great interest, being related to realization of the principle of correspondence between classical and quantum mechanics.

Conventionally, the QC system is the time-inversion invariant system whose dynamics is governed by quantum (or wave) equations and whose classical analog possesses chaotic properties. According to Bohigas, Giannoni and Schmit conjecture \cite{bib:BohGianSchmit84}, statistical properties of such a system may be described with the aid of Random Matrix Theory \cite{bib:GurMullerWeiden98}. The main signatures of QC that result from this theory are the spectrum chaotic appearance and the correlation between spectral lines, the latter being manifested through the peculiar level repulsion and through the Wigner (or close) distribution of inter-frequency (IF) intervals.

In studies on quantum chaos, the scattering billiards of Sinai and Bounimovich type are normally used, in which QC signatures were detected experimentally \cite{bib:Stockman99}. The signatures are also pertinent to other systems, such as scattering billiards with randomly rough boundaries \cite{bib:GanapEremTar07}, volume billiards with random bulk inhomogeneities \cite{bib:GanapEremTar09,bib:GanapTarShost11}, cylindrical billiards with kinks of the lateral boundaries, where the second derivative disappears at some distinct points \cite{bib:Gurevich98,bib:Ganap12}. Besides that, the presence of QC signatures was predicted for quantum/wave systems containing one or more quite small inclusions, each being treated as a~singular perturbation \cite{bib:Shigehara94,bib:ShigeharaCheon96}. Yet, in spite of the fruitfulness of the theoretical method applied in Refs.~\cite{bib:Shigehara94,bib:ShigeharaCheon96} a number of questions were not properly clarified in those papers. In particular, one of the fundamental questions remained unclear, viz., what is the actual physical reason for the quantum or wave system subject to small spatial perturbation to take on the well-defined QC signatures?

\begin{figure}[h]
  \centering
  \scalebox{.6}[.6]{\includegraphics{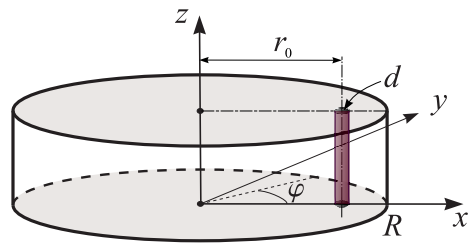}}
  \caption{(Color online) Cylindrical quasioptical cavity resonator with narrow metallic rod assigned to simulate a singular perturbation.
  \label{fig1}}
\end{figure}

In the present paper, the goal is to examine, both theoretically and experimentally, spectral properties of a cylindrical quasioptical cavity resonator of radius $R$ supplied with the metallic rod of small diameter $d\ll\lambda$ ($\lambda$~is the wave length) positioned parallel to the resonator axis at an arbitrary distance $r_0$ from it (see Fig.~\ref{fig1}).
In the presence of the rod the resonator axial symmetry is violated, so the system becomes non-integrable from the viewpoint of classical particle dynamics. As to the dynamics of waves, on introducing the rod the scattering arises between oscillation modes, which, as is shown below, has a substantial impact upon the spectrum statistical properties even if the rod diameter is rather small.

From classical mechanics position, the resonator with a thin rod inside it may be viewed as a singularly perturbed Sinai billiard, in which the trajectory instability results from particle scattering at the rod surface of modulo large negative curvature. For wave systems, the question of fundamental concern is whether or not the inter-mode scattering caused by the perturbation of singular nature could result in the appearance of QC signatures analogous to those characteristic for systems with classical dynamics. In particular, these are the randomly looking spectrum, the presence of correlation between spectral lines, and the Wigner-type distribution of the IF intervals.

In theoretical analysis of this spectral problem we simulate the rod inserted into the resonator by adding the model effective potential to the wave equation (by analogy with quantum mechanics). The form and the magnitude of the potential are meant to be specified based on the spectral measurement data. Such a method is quite transparent for the physical interpretation, as it provides a~way for direct determination of the potential parameters. The resonator spectrum is determined from the Green function equation,
\begin{equation}\label{Wave_eq-gen}
  \left[\Delta+k^2-i/\tau_d-V(\mathbf{r})\right]G(\mathbf{r},\mathbf{r}')=\delta(\mathbf{r}-\mathbf{r}') \ ,
\end{equation}
where $\Delta$ is the three-dimensional Laplacian, ${k^2=(\omega/c)^2}$, and $1/\tau_d$ is the dissipation rate we introduce phenomenologically; function $V(\mathbf{r})$ is the effective potential meant to simulate the metallic rod.

Suppose that E-type oscillations are excited in the resonator, whose electric field is directed along axis $z$. The potential in Eq.~\eqref{Wave_eq-gen}, allowing for its singular nature, can be modeled as
\begin{equation}\label{V(r)-delta}
  V(\mathbf{r})=\Lambda(k,d)\delta(\mathbf{r}_{\perp}-\mathbf{r}_0)\ ,
\end{equation}
where $\Lambda(k,d)$ is the dimensionless parameter dependent on the oscillation frequency and the rod diameter $d$; $\mathbf{r}_{\perp}$~is the radius-vector component perpendicular to axis $z$, $\mathbf{r}_{0}$~is the plain vector that specifies the position of the perturbing rod relative to the resonator main axis.

In Fourier representation, equation Eq.~\eqref{Wave_eq-gen} takes on the form of an infinite set of coupled equations for Green function matrix elements,
\begin{align}\label{Main_eq-mode}
  (k^2  - \varkappa_{\bm{\mu}}^2 - i/\tau _d -\mathcal{V}_{\bm{\mu }}) & G_{{\bm{\mu} \bm{\mu }}'} \notag\\ & -
  \sum\limits_{{\bm{\nu }} \ne {\bm{\mu }}}^{} {\mathcal{U}_{{\bm{\mu
  \bm{\nu} }}} } G_{{\bm{\nu \bm{\mu} }}'}  = \delta _{{\bm{\mu
  \bm{\mu} }}'}\ .
\end{align}
Here, $\bm{\mu}=(l,n,q)$ is the vectorial mode index conjugate to coordinate vector $\mathbf{r}=(r,\varphi,z)$, $\varkappa _{\bm{\mu }}^2$ is the (taken with ``minus'' sign) eigenvalue of the Laplace operator corresponding to the $\bm{\mu}$-th spatial mode of the empty resonator. We will refer to this quantity as the ``unperturbed mode energy''. Quantities $\mathcal{U}_{{\bm{\mu\bm{\nu}}}}$ and $\mathcal{V}_{\bm{\mu}}\equiv \mathcal{U}_{{\bm{\mu\bm{\mu}}}}$ are the mode matrix elements of the potential Eq.~\eqref{V(r)-delta} taken between the eigenfunctions of the unperturbed resonator,
\begin{align}
 \label{Umunu}
  \mathcal{U}_{\bm{\mu}\bm{\nu}} & =\int_\Omega d\mathbf{r}
  \bra{\mathbf{r};\bm{\mu}}V(\mathbf{r})\ket{\mathbf{r};\bm{\nu}}\ .
\end{align}

In paper Ref.~\cite{bib:GanapEremTar07} it was shown that equations Eqs.~\eqref{Main_eq-mode} can be solved for the entire set of the Green function matrix elements using the special functional method of mode separation. Within this method, the non-diagonal elements of Green matrix $\{G_{\bm{\nu\mu}}\}$ are linearly expressed through the diagonal ones, whose equations are strictly decoupled and then solved to
\begin{equation}\label{Gmumu}
  G_{\bm{\mu\mu}}  = \left( k^2-\varkappa_{\bm{\mu }}^2-i/\tau_d-\mathcal{V}_{\bm{\mu}}-\mathcal{T}_{\bm{\mu}}\right)^{-1}\ .
\end{equation}
Owing to the linear relationship between ``skew'' and diagonal Green matrix elements, the poles of exactly Green functions Eq.~\eqref{Gmumu} determine the resonator spectrum.

In expression Eq.~\eqref{Gmumu}, along with scalar potential $\mathcal{V}_{\bm{\mu}}$ describing the impact on the spectrum of scattering events with no change in the mode index (the so-called \emph{intra-mode} scattering), the potential $\mathcal{T}_{\bm{\mu}}$ is also present, which possesses the operator structure. This potential is nothing else but the regularized \emph{T}-matrix, which allows for the inter-mode scattering and the associated mode intermixing. The exact form of the \emph{T}-potential is
\begin{equation}\label{T-potential}
  \mathcal{T}_{\bm{\mu}}=\hat{\mathbf{P}}_{\bm{\mu}}\hat{\mathcal{U}}\big(1-\hat{\mathsf{R}}\big)^{- 1} \!\hat{\mathsf{R}}\,\hat{\mathbf{P}}_{\bm{\mu}}\ ,
\end{equation}
where $\hat{\mathbf{P}}_{\bm{\mu}}$ is the projection operator the action of which is to assign pre-specified value $\bm{\mu}$ to the nearest mode index of any operator standing next to it, both to the left and to the right; $\hat{\mathcal{U}}$ is the null-diagonal matrix whose non-vanishing elements are inter-mode potentials $\mathcal{U}_{\bm{\mu}\bm{\nu}}$ with ${\bm{\mu}\neq\bm{\nu}}$, ${\hat{\mathsf{R}}=\hat{G}^{(V)}\hat{\mathcal{U}}}$ is the mode-mixing operator which includes the ``seed'' Green operator $\hat{G}^{(V)}$ with diagonal mode matrix consisting of the entire set of trial Green functions. These functions are the solution to equation set \eqref{Main_eq-mode} with all the inter-mode potentials temporarily set to zero, viz.
\begin{equation}\label{Gmu(V)}
  G_{\bm{\mu}}^{(V)}  = \left(k^2-\varkappa_{\bm{\mu }}^2-\mathcal{V}_{\bm{\mu}}-i/\tau _d \right)^{ - 1}\ .
\end{equation}
The functions Eq.~\eqref{Gmu(V)} may be regarded as constituting the basis for the development of perturbation theory with regard to the inter-mode potentials only, the intra-mode scattering being accounted for in full.

Although the spectrum of the resonator with the inserted rod is completely determined by propagators Eq.~\eqref{Gmumu}, the practical usage of formula Eq.~\eqref{T-potential}, in view of its complicated functional structure, is difficult in the general case. Yet, possessing the exact expression for the \emph{T}-potential that takes the account of the inter-mode scattering, one can analyze in detail the limiting cases of weak and strong mode correlation, bearing in mind small or large (as compared with unity) norm of mode-mixing operator $\hat{\mathsf{R}}$. To calculate this norm one has to use the exact form of potentials Eq.~\eqref{Umunu}. Within the model Eq.~\eqref{V(r)-delta}, these potentials are easily calculated and become
\begin{widetext}
\begin{subequations}\label{Intra+Inter_potentials}
\begin{align}
\label{Vmu-fin}
 \mathcal{V}_{\bm{\mu}} & =\frac{1}{\pi R^2} \Lambda(k,d)C^2_{l_{\bm{\mu}}n_{\bm{\mu}}}
  J^2_{|n_{\bm{\mu}}|}\left({\gamma_{l_{\bm{\mu}}}^{(|n_{\bm{\mu}}|)} r_0\big/R}\right)\ ,
  \\[.5\baselineskip]
 \mathcal{U}_{\bm{\mu}\bm{\nu}} & = \delta_{q_{\bm{\mu}}q_{\bm{\nu}}}
  \frac{\Lambda(k,d)}{\pi R^2} C_{l_{\bm{\mu}}n_{\bm{\mu}}}C_{l_{\bm{\nu}}n_{\bm{\nu}}}
J_{|n_{\bm{\mu}}|}\left({\gamma_{l_{\bm{\mu}}}^{(|n_{\bm{\mu}}|)}r_0\big/R}\right)
  J_{|n_{\bm{\nu}}|}\left({\gamma_{l_{\bm{\nu}}}^{(|n_{\bm{\nu}}|)}r_0\big/R}\right)\ ,
 \hspace{1cm}
\label{Umunu-fin-old}
\end{align}
\end{subequations}
%
where numerical coefficients $C_{ln}=J_{|n|+1}^{-1}\left({\gamma _l^{(|n|)}}\right)$ originate from eigenfunction normalization. With potentials Eqs.~\eqref{Intra+Inter_potentials}, the operator $\hat{\mathsf{R}}$ square norm is calculated to
\begin{align}\label{R-norm_explicit}
  \|\hat{\mathsf R}\|^2=\left[\frac{\Lambda(k,d)}{\pi R^2}\right]^2
  \max_{\bm{\mu}} & \Bigg[\sum_{l_{\bm{\nu}},n_{\bm{\nu}}}
  \delta_{q_{\bm{\nu}}q_{\bm{\mu}}}\frac{C_{l_{\bm{\nu}}n_{\bm{\nu}}}C_{l_{\bm{\mu}}n_{\bm{\mu}}}}
  {k^2-\varkappa_{\bm{\nu}}^2+i/\tau _d-\mathcal{V}_{\bm{\nu}}}
  J_{|n_{\bm{\nu}}|}\left({\gamma_{l_{\bm{\nu}}}^{(|n_{\bm{\nu}}|)}r_0\big/R}\right)
  J_{|n_{\bm{\mu}}|}\left({\gamma_{l_{\bm{\mu}}}^{(|n_{\bm{\mu}}|)}r_0\big/R}\right)\Bigg]\notag\\
  \times &
  \Bigg[\sum_{l_{\bm{\nu}'},n_{\bm{\nu}'}}
  \delta_{q_{\bm{\nu}'}q_{\bm{\mu}}}\frac{C_{l_{\bm{\nu}'}n_{\bm{\nu}'}}C_{l_{\bm{\mu}}n_{\bm{\mu}}}}
  {k^2-\varkappa_{\bm{\nu}'}^2-i/\tau _d-\mathcal{V}_{\bm{\nu}'}}
  J_{|n_{\bm{\nu}'}|}\left({\gamma_{l_{\bm{\nu}'}}^{(|n_{\bm{\nu}'}|)}r_0\big/R}\right)
  J_{|n_{\bm{\mu}}|}\left({\gamma_{l_{\bm{\mu}}}^{(|n_{\bm{\mu}}|)}r_0\big/R}\right)\Bigg]\ .
\end{align}
\end{widetext}
This expression enables the order-of-magnitude estimation of $\|\hat{\mathsf R}\|$ in different limiting cases, and in this way makes it possible to specify the degree of the mode intermixing, which is responsible for the spectrum chaotic property.

The most interesting for us is the limiting case where the intra-mode scattering is weak and does not noticeably affect the mode correlation. The condition for this is the fulfilment of inequality
\begin{equation}\label{Weak_intramode}
  \Lambda\left(k,d\right) \ll (kR)^2\ ,
\end{equation}
where the quantity on the right-hand side is large as compared to unity. On condition Eq.~\eqref{Weak_intramode}, the intra-mode potentials in expression Eq.~\eqref{R-norm_explicit} may be safely omitted, thus giving the following estimation for the mode-mixing operator norm,
\begin{equation}\label{R_norm-estim}
  \|\hat{\mathsf R}\|^2\sim\Lambda^2\left(k,d\right)\ .
\end{equation}
Based on this estimate one can infer that under weak intra-mode scattering, the level of mode intermixing, and therefore the degree of the state of chaos in the spectrum, may differ essentially, depending on whether the coupling constant entering Eq.~\eqref{V(r)-delta} is small or large as compared to unity. If $\|\hat{\mathsf R}\|\ll 1$, the modes are correlated relatively weakly and in fact are clusterized. Conversely, if ${\|\hat{\mathsf R}\|\gg 1}$ the mode spectrum is to be considered as correlated ergodically.

To elucidate the questions that arise in the analysis of spectrum chaotic properties it is necessary to specify the model coupling constant $\Lambda$, i.\,e., its value and the behavior as a function of governing parameters, which are the oscillation frequency and the diameter of the inserted rod. Since the exact form of the model potential is indeterminate, for the valid choice of the coupling constant in Eq.~\eqref{V(r)-delta} we have performed a~set of experiments with a quasioptical cavity resonator, where the singular scatterer was realized in the form of the asymmetrically positioned thin metal rod (see Fig.~\ref{fig1}). A resonator with the inserted circular rod is similar to the classical Sinai billiard, which is known to be a nonintegrable system where the trajectory instability results from particle scattering at the rod lateral surface of negative curvature. In wave systems, the trajectory concept is inappropriate, so in this case an issue arises whether the wave scattering by small-diameter cylindric insert could result in the appearance of main signatures of the quantum chaos.

To answer this query, in the resonator of 130~mm diameter and 15~mm in height the oscillations were excited by means of diffraction antenna over the frequency range of $26 \div 39$~GHz. The antenna was the round hole 2~mm in diameter in a thin copper membrane (0.1~mm thick) that shuts the outlet from the rectangular launching waveguide. With such an antenna, both H and E type oscillations were excited. In the former case, the wide side of the waveguide was directed along the resonator vertical axis whereas in the latter case it was oriented in the perpendicular plane. To record the measurement data, the wide-band gage, which enables determining the signal passing through the resonator, was used. The gage allowed one, during the comparatively short period of time (of about 40~sec.), not only to computer-record the resonator spectrum in the required frequency range but also to determine the resonance frequency and the width of each of the multiple (a few hundreds) spectral lines. The calibrator built into the special computer programm have made it possible to ensure quite high accuracy of spectral line measurements. The relative error in determining the spectral line frequencies and their widths was $10^{-6}$ and $10^{-4}$, respectively. It was also possible to vary the magnitude of the singular perturbation using the rods of different diameters, thereby adjusting the inter-mode scattering level.

The measured spectra of E and H oscillations were found to be of the same qualitative character. Both of them
\begin{figure}[h]
  \centering
  \scalebox{.45}[.45]{\includegraphics{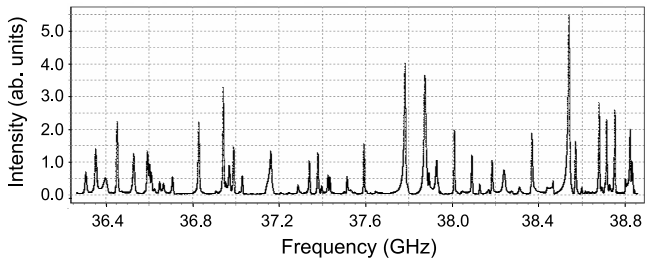}}
  \caption{The fragment of the resonator frequency spectrum.
  \label{fig2}}
\end{figure}
consist of spectral lines that seem to be randomly distributed over the frequency axis (see Fig.~\ref{fig2}). Yet, the quality factors differ essentially both for E and H oscillations, and for the oscillations of the same type. Note that the H oscillations Q-factors were appreciably greater as compared to the Q-factors for E-type oscillations.

To specify the degree of the resonance mode coupling to the additional object in the resonator we employed the following technique. We selected a characteristic line within the spectrum, and for this line the frequency shift was measured when inserting the rods of different geometric parameters. When estimating the inter-mode scattering rate, the coupling of different modes with the perturbation was reckoned to be the same, to an accuracy of an order of magnitude. The measurements have revealed that the resonance frequency shift is, firstly, proportional to the rod diameter, $|\Delta\omega_{\bm{\mu}}|\propto d$. Secondly, this shift is of relatively low value, $|\Delta\omega_{\bm{\mu}}|/\omega_{\bm{\mu}}\ll 1$. To estimate the coupling constant from this shift, we have used the asymptotic formula for the correction to the mode energy. The formula follows from Eq.~\eqref{T-potential} in the limit of weak inter-mode scattering and is given by
\begin{align}\label{Mode_shift}
  \Delta\varkappa_{\bm{\mu }}^2= & \frac{\Lambda^2(\varkappa_{\bm{\mu}},d)}{(\pi R^2)^2}
  \left[C_{l_{\bm{\mu}}n_{\bm{\mu}}}
  J_{|n_{\bm{\mu}}|}\left({\gamma_{l_{\bm{\mu}}}^{(|n_{\bm{\mu}}|)}r_0\big/R}\right)\right]^2\times
  \notag\\
 & \sum_{{\bm{\nu}}\neq{\bm{\mu}}}\delta_{q_{\bm{\nu}}q_{\bm{\mu}}}
  \left[C_{l_{\bm{\nu}}n_{\bm{\nu}}}
  J_{|n_{\bm{\nu}}|}\left({\gamma_{l_{\bm{\nu}}}^{(|n_{\bm{\nu}}|)}r_0\big/R}\right)\right]^2
  \times
  \notag\\
 & \frac{\varkappa_{\bm{\mu}}^2-\varkappa_{\bm{\nu}}^2}
  {(\varkappa_{\bm{\mu}}^2-\varkappa_{\bm{\nu}}^2)^2+1/\tau^2_d}\ .
\end{align}
According to this expression, the coupling constant is estimated as $\Lambda\sim\sqrt{|\Delta\omega_{\bm{\mu}}|/\omega_{\bm{\mu}}}\ll 1$, which is entirely consistent with the approximations made in our theory.

From Eq.~\eqref{Mode_shift} it can be easily seen that the shift of the resonance peak (specifically, of the $\bm{\mu}$-th one) is determined substantially by the neighboring resonance frequencies. This is evidenced by the presence of the sum over mode indices $\bm{\nu}\neq\bm{\mu}$, whose terms contain resonance multipliers. The resonance frequencies are thus made considerably correlated due to the inhomogeneity-induced inter-mode scattering, and this is clearly indicative of the chaos in the spectral line distribution.

We have analyzed the spectrum statistical properties by plotting the histograms of IF
\begin{figure}[h]
  \centering
  \scalebox{.5}[.45]{\includegraphics{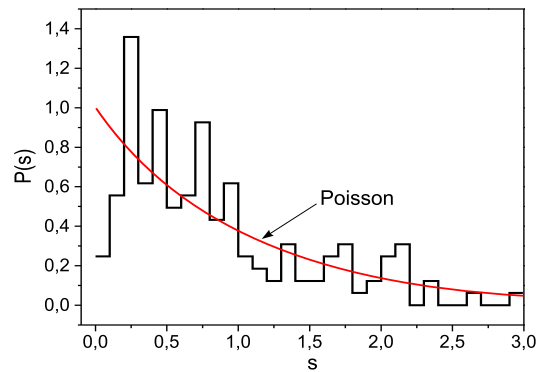}}
  \caption{(Color online) The IF interval histogram for the unperturbed resonator; $s$ is the normalized IF interval length. For comparison, the solid line indicates the Poisson distribution.
  \label{fig3}}
\end{figure}
intervals for the empty resonator and for resonators with different perturbing rods. For the resonator with no rod, the histogram depicts virtually the Poisson IF interval distribution
(see Fig.~\ref{fig3}), both for E and for H oscillations. This testifies manifestly the spectral lines statistical independence.

\begin{figure}[t]
  \centering
  \scalebox{.7}[.7]{\includegraphics{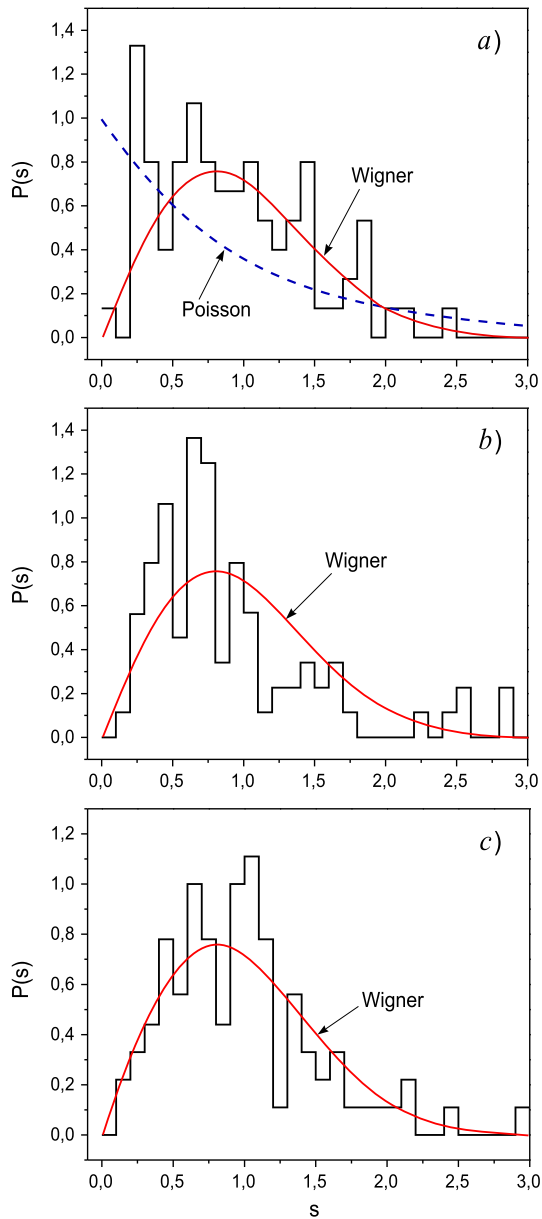}}
  \vspace{-2mm}
  \caption{(Color online) The IF interval histograms for the resonators perturbed by the rods of different diameters: a)~$d=0.37$~mm, $\Lambda=0.022$; b) $d= 0.7$~mm, $\Lambda=0.032$; c)~$d=1.5$~mm, $\Lambda=0.045$ (the error of $d$ is $\approx 10^{-3}$~mm). The rod offset is $r_0=38$~mm.
  \label{fig4}}
\end{figure}
In Figure~\ref{fig4}, the histograms are shown for resonators with the inserted rods of different diameters. It is easily seen that the singular perturbation leads to the significant changes in the spectrum statistical properties, thus making them largely dependent upon the degree of the inter-mode scattering. With a growing diameter of the rod, the coupling parameter in potential Eq.~\eqref{V(r)-delta} increases progressively, which results in the appreciable re-distribution of resonance frequencies and in the gradual Wignerization of the IF interval spectrum.

Even for relatively small diameters of the rod (${d\approx 0.4}$~mm), the histogram for E-type oscillations (Fig.~\ref{fig4}\emph{a}) exhibits an appreciable difference of IF interval distribution from the Poissonian form. Such a distribution may be thought as an intermediate one between Poisson and Wigner distributions. As the rod diameter increases ($d>0.6$~mm), the histogram envelope assumes the form
\begin{figure}[t]
  \centering
 \vspace{4mm}
  \scalebox{.5}[.5]{\includegraphics{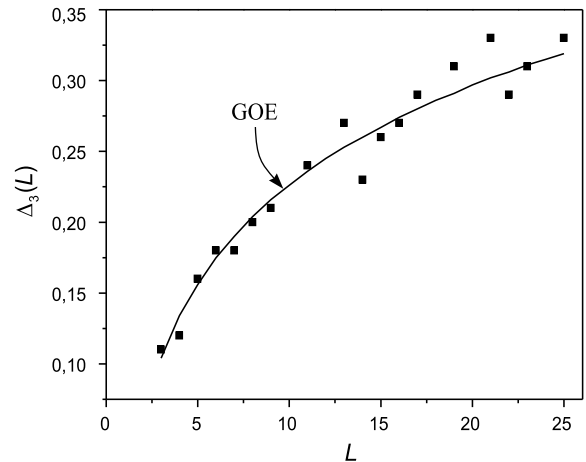}}
  \caption{Experimental and theoretical data for spectral rigidity of singularly perturbed resonator with $\Lambda=0.045$ and ${d  = 1.5}$~mm. $L$ is the normalized length of the sequence of measured resonance frequencies \cite{bib:Stockman99}, the solid line corresponds to GOE statistics.
  \label{fig5}}
\end{figure}
ever more close to the Wigner curve. This is an evidence of the QC signature appearance in the resonator with a relatively strong, yet singular in space, perturbation. The results for spectral rigidity in Fig.~\ref{fig5} also indicate the appearance of quantum chaos in the spectrum of singularly perturbed resonator. The measurements were made in accordance with the methodology described in  Ref.~\cite{bib:Reichl04}.

Additionally, we have conducted the measurements of IF interval correlations at different levels of singular perturbation of the resonator in question. The results are in qualitative consistence with the data extracted from the above-shown histograms. Specifically, the correlation between the nearest levels of the rod-free resonator appears to be very small, ${C(1)=0.003}$. This matches almost ideally the value corresponding to the independent lines of the perturbation-free circular resonator. For the resonator equipped  with the rod of diameter $d=1.5$~mm, whose histograms show considerable chaos signatures, the correlation coefficient $C(1)=-0.268$. This agrees fairly well with theoretical estimates of coefficient $C(1)$ for systems pertaining to the GOE universality class~\cite{bib:BohGianSchmit84,bib:Elutin88}.

Our experimental data likewise have suggested that factor $\Lambda$ has much larger values in the case of E oscillations than for oscillations of H type. Besides, it results from the estimates we have conducted that the intra-mode scattering level is much lower as against that of the inter-mode scattering. This suggests that at a sufficiently high level of singular perturbation ($kd>1$) the inter-mode scattering is prevailing. As is seen from Eq.~\eqref{Mode_shift}, this scattering mixes the resonance system states, thereby serving as an origin of the quantum chaos.



\begin{thebibliography}{cc}

\bibitem{bib:Stockman99}
H.-J. St\"ockman, \textit{Quantum Chaos: An Introduction} (Cambridge University Press, Cambridge, 1999).

\bibitem{bib:BohGianSchmit84}
O. Bohigas, M.\,J. Giannoni, C. Schmit,  \prl \textbf{52}, 1 (1984).

\bibitem{bib:GurMullerWeiden98}
T. Guhr, A. M\"uller-Groeling, and H. Weidenm\"uller,  Phys. Rep. \textbf{299}, 189 (1998).

\bibitem{bib:GanapEremTar07}
E.\,M. Ganapolskii, Z.\,E. Eremenko, and Yu.\,V. Tarasov, \pre \textbf{75}, 026212 (2007).

\bibitem{bib:GanapEremTar09}
E.\,M. Ganapolskii, Z.\,E. Eremenko, and Yu.\,V. Tarasov, \pre \textbf{79}, 041136 (2009).

\bibitem{bib:GanapTarShost11}
E.\,M. Ganapolskii, Yu.\,V. Tarasov, and L.\,D. Shostenko, \pre \textbf{84}, 026209 (2011).

\bibitem{bib:Gurevich98}
B.M.Gurevich, \textit{Ergodichnost' dinamicheskich system}. In book: \textit{Physical encyclopedia}, v.~5, ed.~A.\,M.~Prokhorov (Ros. Encyclopedia, Moskow, 1998, in Russian).

\bibitem{bib:Ganap12}
E.\,M. Ganapolskii, JETP Lett. \textbf{96}, 456 (2012).

\bibitem{bib:Shigehara94}
T. Shigehara, \pre \textbf{50}, 4357 (1994).

\bibitem{bib:ShigeharaCheon96}
T. Shigehara and T. Cheon, \pre \textbf{54}, 1321 (1996).

\bibitem{bib:Reichl04}
L.\,E. Reichl, \textit{The Transition to Chaos: Conservative Classical Systems and Quantum Manifestations} (Springer-Verlag,  New York, 2004).

\bibitem{bib:Elutin88}
P.\,V. Elyutin, Sov. Phys. Usp. \textbf{31}, 597 (1988).

\end{thebibliography}
\end{document}